  \documentclass[graphicx,amssymb,amsmath,enumerate,11pt]{report}
  \usepackage{fancyhdr}
  \fancyheadoffset[]{0.1cm}

  \usepackage{amsmath}
  \usepackage{epsfig}
  \usepackage{axodraw}
  \usepackage{url}
  \usepackage{footnote}
  \makesavenoteenv{tabular}

\newcommand{\neff}{N_{\textrm{eff}}}
\newcommand{\summnu}{\sum m_\nu}
\newcommand{\yhe}{Y_{\textrm{he}}}
\newcommand{\ceff}{c_{\textrm{eff}}^2}
\newcommand{\cvis}{c_{\textrm{vis}}^2}

\begin{document}





\begin{center}
{\Large \bf Cosmic dark radiation and neutrinos}\\
{\it Maria Archidiacono~\footnote{Department of Physics and Astronomy University of Aarhus, DK-8000 Aarhus C, Denmark}$^*$, Elena Giusarma~\footnote{IFIC, Universidad de Valencia-CSIC, 46071, Valencia, Spain}, Steen Hannestad~$^1$, Olga Mena~$^2$}
\end{center}
\let\thefootnote\relax\footnote{$^*$archi@phys.au.dk}

\vskip 0.5cm

\begin{center}
{\bf Abstract}
\end{center}
New measurements of the cosmic microwave background (CMB) by the Planck mission have greatly increased our knowledge about the Universe. Dark radiation, a weakly interacting component of radiation, is one of the important
ingredients in our cosmological model which is testable by Planck and other observational probes. At the moment the possible existence of dark radiation is an unsolved question. 
For instance, the discrepancy between the value of the Hubble constant, $H_0$, inferred from the Planck data and local measurements of $H_0$ can to some extent be alleviated by enlarging the minimal $\Lambda$CDM model to include additional relativistic degrees of freedom.
From a fundamental physics point of view dark radiation is no less interesting. Indeed, it could well be one of the most accessible windows to physics beyond the standard model. An example of this is that sterile neutrinos, hinted at in terrestrial oscillation experiments, might also be a source of dark radiation, and cosmological observations can therefore be used to test specific particle physics models.
Here we review the most recent cosmological results including a complete investigation of the dark radiation sector in order to provide an overview of models that are still compatible with new cosmological observations. Furthermore we update the cosmological constraints on neutrino physics and dark radiation properties focussing on tensions between data sets and degeneracies among parameters that can degrade our information or mimic the existence of extra species.

\vskip 1cm


\section{Introduction}

The connection between cosmological observations and neutrino physics
is one of the most interesting and hot topics in astroparticle physics.

Earth based experiments have demonstrated that neutrinos oscillate and
therefore have mass (see e.g.\ \cite{Tortola:2012te} for a recent treatment). However oscillation experiments are not sensitive to
the absolute neutrino mass scale, only the squared mass differences, $\Delta m^2$. 
Furthermore, the sign is known for only one of the two mass differences, namely $\Delta m_{12}^2$, because of matter effects in the Sun.
$\Delta m_{23}^2$ is currently only measured via vacuum oscillations which depends only on $|\Delta m_{23}^2|$.
Even for standard model neutrinos there are therefore important unresolved questions which have a significant impact on cosmology. Not only is the absolute 
mass scale not known, even the hierarchy between masses is unknown.
In any case the two measured mass squared differences imply that at least two
neutrinos are very non-relativistic today (see e.g. Ref. \cite{Fogli:2012ua} for a recent overview).

Unlike neutrino oscillation experiments, cosmology probes the sum of the neutrino masses (see e.g.\ \cite{Hannestad:2010kz,Wong:2011ip}) because it is sensitive 
primarily to the current neutrino contribution to the matter density. At the moment cosmology provides a stronger bound on the neutrino mass than 
laboratory bounds from e.g.\ beta decay, although the KATRIN experiment is set to improve the sensitivity to $\summnu$ to about 0.6 eV \cite{Osipowicz:2001sq}.

The tightest 95\% c.l. upper limits to date are $\summnu<0.15$~eV \cite{Riemer-Sorensen:2013jsa} and $\summnu<0.23$~eV \cite{Giusarma:2013wsa} from different combinations of data sets and different analyses.
This astounding accuracy is possible because neutrinos leave key signatures
through their free-streaming nature
in several cosmological data sets: The temperature-anisotropy power spectrum of the Cosmic Microwave Background (see Section \ref{sec:effects}) and
the power spectrum of matter fluctuations, which is
one of the basic products of galaxy redshift surveys (see Ref. \cite{Lesgourgues:2012uu}).
However, it should be stressed that cosmological constraints are highly model dependent and, following the Bayesian method,
theoretical assumptions have a strong impact on the results and can lead to erroneous conclusions. For instance in Refs. \cite{GonzalezGarcia:2010un, Smith:2011ab} the assumption about spatial flatness is relaxed, testing therefore the impact of a non zero curvature in the neutrino mass bound. It is also well-known that the bound on the neutrino mass is sensitive to assumptions about the dark energy equation of state \cite{Hannestad:2005gj}.

In the standard model there are exactly three neutrino mass eigenstates, $(\nu_1\nu_2,\nu_3)$, corresponding to the three flavour eigenstates ($\nu_{e}$, $\nu_{\mu}$, $\nu_{\tau}$) of the weak interaction.

This has been confirmed by precision electroweak measurements at the
$Z^0$-resonance by the LEP experiment. The invisible decay width of $Z_0$ corresponds to 
$N_{\nu}\:=\:2.9840 \pm 0.0082$ \cite{Abbiendi:2000hu},
consistent within $\sim 2\,\sigma$ with the known
three families of the SM.

In cosmology the energy density contribution of one ($\neff=1$) fully thermalised neutrino plus anti-neutrino below the $e^+e^-$ annihilation scale of $T \sim 0.2$ MeV is at lowest order given by $\rho_\nu = \frac{7}{8} \left(\frac{4}{11}\right)^{4/3} \rho_\gamma$. However, a more precise calculation which takes into account finite temperature effects on the photon propagator and incomplete neutrino decoupling during $e^+e^-$ annihilation leads to a standard model prediction of $\neff\:=\:3.046$ (see e.g.\ \cite{Mangano:2005cc}). This is not because there is a non-integer number of neutrino species but simply comes from the definition of $\neff$.

In the last few years the WMAP satellite as well as the high multipole CMB experiments
Atacama Cosmology Telescope (ACT) and South Pole Telescope (SPT)
provided some hints for a non standard value of the effective number of relativistic degrees of freedom $\neff$, pointing towards the existence of an extra dark component of the radiation content of the Universe,
coined dark radiation.

A variation in $\neff$ affects
both the amplitude and the
shape
of the Cosmic Microwave Background
temperature anisotropy power spectrum (see Section \ref{sec:effects}).
Nevertheless the new data releases of these two experiments (see Ref. \cite{Sievers:2013ica} for ACT and Ref. \cite{Hou:2012xq} for SPT) seem to disagree in their conclusions on this topic \cite{Archidiacono:2013lva}:
in combination with data from the last data release of the Wilkinson Microwave Anisotropy Probe satellite (WMAP 9 year),
SPT data lead to an evidence of an extra dark radiation component ($\neff\:=\:3.93\pm0.68$), while ACT data prefer a standard value of $\neff$ ($\neff\:=\:2.74 \pm 0.47$). The inclusion of external data sets (Baryonic Acoustic Oscillation \cite{bao} and Hubble Space Telescope measurements \cite{Riess:2011yx}) partially reconciles the two experiments in the framework of a $\Lambda$CDM model with additional relativistic species.

The recently released Planck data have strongly confirmed the standard $\Lambda$CDM model. The results have provided the most precise constraints ever on the six "vanilla" cosmological parameters \cite{Ade:2013zuv} by measuring the Cosmic Microwave Background temperature power spectrum up to the seventh acoustic peak \cite{Planck:2013kta} with nine frequency channels (100, 143, 217 GHz are the three frequency channels involved in the cosmological analysis). Concerning dark radiation, Planck results point towards a standard value of $\neff$ ($\neff\:=\:3.36^{+0.68}_{-0.64}$ at 95\% c.l. using Planck data combined with WMAP 9 year polarization measurements and high multipole CMB experiments, both ACT and SPT). However the $\sim2.5\sigma$ tension among Planck and HST measurements of the Hubble constant value can be solved, for instance, by extending the $\Lambda$CDM model to account for a non vanishing $\Delta \neff$ ($\neff\:=\:3.62^{+0.50}_{-0.48}$ at 95\% c.l. using Planck+WP+highL plus a prior on the Hubble constant from the Hubble Space Telescope measurements \cite{Riess:2011yx}).

In this review, after explaining the effects of $\neff$ on CMB power spectrum (Section \ref{sec:effects}), in Section \ref{sec:models} we list the different dark radiation models with their state of art constraints on the effective number of relativistic degrees of freedom. Section \ref{sec:method} illustrates the method and the data sets we use here in order to constrain the neutrino parameters we are interested in (number of species and masses). The results of our analyses are reported in Section \ref{sec:results}. Finally in Section \ref{sec:conclusions} we discuss our conclusions in light of the former considerations.

\subsection{$\neff$ effects on cosmological observables}
\label{sec:effects}

The total radiation content of the Universe below the $e^+e^-$ annihilation temperature can be parametrized as follows:
\begin{equation}
\varrho_{r}=\left[1+\frac{7}{8}\left(\frac{4}{11}\right)^{4/3}\neff\right]\varrho_{\gamma},
\label{eq:rhor}
\end{equation}
where $\rho_\gamma$ is the energy density of photons, $7/8$ is the multiplying factor for each fermionic degree of freedom and $(4/11)^{1/3}$ is the photon neutrino temperature ratio.
Finally the parameter $\neff$ can account for neutrinos and for any extra relativistic degrees of freedom, namely particles still relativistic at decoupling:
\begin{displaymath}
\neff=3.046+\Delta\neff.
\end{displaymath}

\begin{figure*}
\begin{center}
\includegraphics[width=13cm]{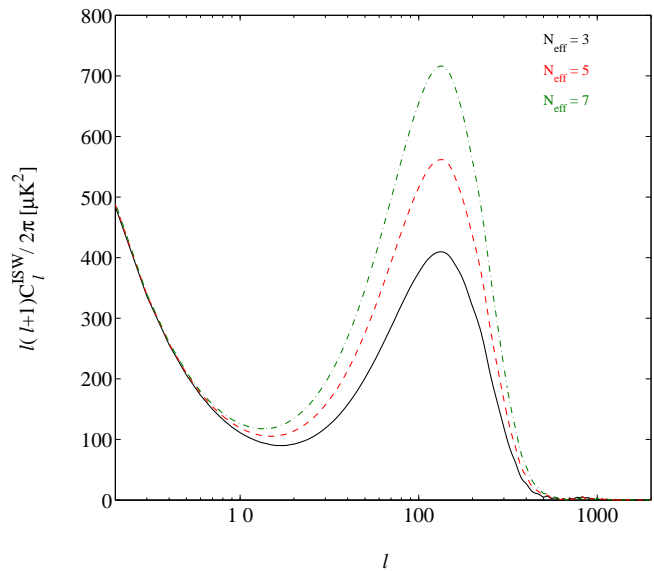}
\caption{ISW contribution to the CMB temperature power spectrum. The raise at $\ell<30$ is due to the late Integrated Sachs Wolfe, while the peak around $\ell\sim200$ is the early Integrated Sachs Wolfe effect. The cosmological model is the $\Lambda$CDM with $\neff$ equals to 3 (black solid line), 5 (red dashed line) and 7 (green dot-dashed line).}
\label{fig:isw}
\end{center}
\end{figure*}

Varying $\neff$ changes the time of the matter radiation equivalence: a higher radiation content due to the presence of additional relativistic species leads to a delay in $z_{\rm eq}$:
\begin{equation}
1+z_{\rm eq}=\frac{\Omega_m}{\Omega_r}=\frac{\Omega_mh^2}{\Omega_\gamma h^2}\frac{1}{\left(1+0.2271\neff\right)},
\label{eq:zeq}
\end{equation}
where $\Omega_m$ is the matter density, $\Omega_r$ is the radiation density, $\Omega_\gamma$ is the photon density, $h$ is defined as $H_{0}=100h$ km/s/Mpc and in the last equality we have used equation (\ref{eq:rhor}).
As a consequence at the time of decoupling radiation is still a subdominant component and the gravitational potential is still slowly decreasing. This shows up as an enhancement of the early Integrated Sachs Wolfe (ISW) effect that increases the CMB perturbation peaks at $\ell\sim200$, i.e.\ around the first acoustic peak as. This effect is demonstrated is in Figure \ref{fig:isw}.

\begin{figure*}
\begin{center}
\includegraphics[width=15cm]{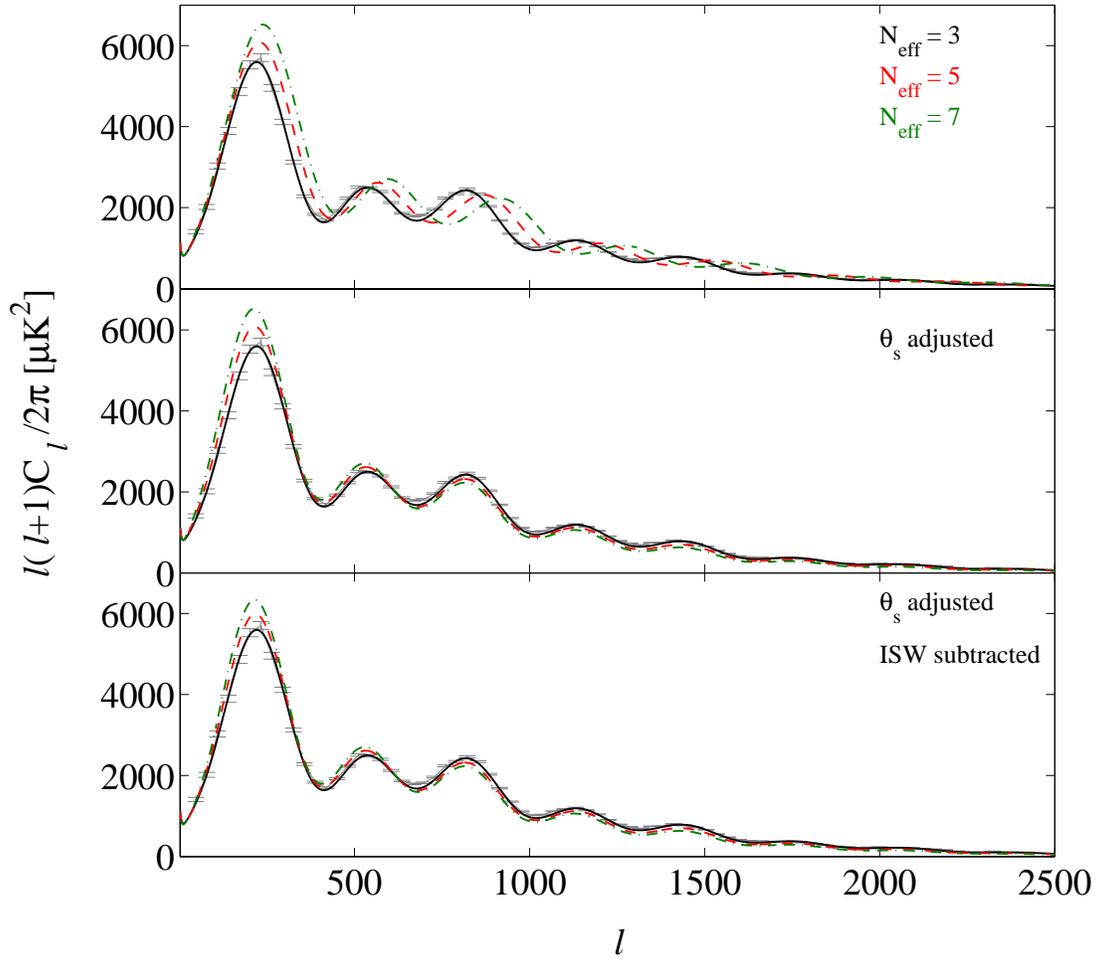}
\caption{CMB temperature power spectrum. The model and the legend are the same as in Figure \ref{fig:isw}, the grey error bars correspond to Planck data.
{\it Top panel}: the total CMB temperature power spectrum.
{\it Middle panel}: the $\ell$ axis has been rescaled by a factor $\theta_{\rm s}(\neff)/\theta_{\rm s}(\neff=3)$.
{\it Bottom panel}: the ISW contribution has been subtracted.}
\label{fig:cls}
\end{center}
\end{figure*}

In \cite{Hou:2011ec} the authors stress that the most important effect of changing $\neff$ is located at high $\ell>600$ and is not related to the early ISW effect. Instead the main effect related to a variation of the number of relativistic species at decoupling is that it alters the expansion rate, $H$, around the epoch of last scattering.
The extra dark radiation component, arising from a value of $\neff$ greater than the standard $3.046$, contributes to the expansion rate via its energy density $\Omega_{\rm DR}$:
\begin{displaymath}
\frac{H^2}{H_0^2}=\frac{\Omega_m}{a^3}+\Omega_\Lambda+\frac{\Omega_\gamma}{a^4}+\frac{\Omega_\nu}{a^4}+\frac{\Omega_{\rm DR}}{a^4}.
\end{displaymath}
If $\neff$ increases, $H$ increases as well.
Furthermore, the delay in matter radiation equality which causes the early ISW also modifies the baryon to photon density ratio:
\begin{displaymath}
R_{\rm eq} = \frac{3 \rho_b}{4 \rho_\gamma} \Big|_{a_{\rm eq}},
\end{displaymath}
and therefore the sound speed
\begin{displaymath}
c_{\rm s}=\frac{1}{\sqrt{3(1+R_{\rm eq})}}.
\end{displaymath}
The size of the comoving sound horizon $r_{\rm s}$ is given by
\begin{displaymath}
r_{\rm s} = \int_0^{\tau'} d\tau c_{\rm s} (\tau) = \int_0^a \frac{da}{a^2 H} c_{\rm s}(a),
\end{displaymath}
and is proportional to the inverse of the expansion rate $r_{\rm s}\propto1/H$,
when $\neff$ increases, $r_{\rm s}$ decreases.
The consequence is a reduction in the angular scale of the acoustic peaks
$\theta_{\rm s}=r_{\rm s}/D_{\rm A}$, where $D_{\rm A}$ is the angular diameter distance.
The overall effect on the CMB power spectrum is a horizontal shift of the peak positions towards higher multipoles. In the middle panel of Figure \ref{fig:cls} the total temperature power spectrum (upper panel) is corrected for this effect: The $\ell$ axis is rescaled by a constant factor $\theta_{\rm s}(\neff)/\theta_{\rm s}(\neff=3)$ in order to account for the peak shift due to the increase in $\neff$. Effectively it amounts to having the same sound horizon for all the models.
Considering that $\theta_{\rm s}$ is the most well constrained quantity by CMB measures, this is the dominant effect of a varying $\neff$ on the CMB power spectrum.

Besides the horizontal shift there is also a vertical shift that affects the amplitude of the peaks at high multipoles where the ISW effect is negligible. Comparing Figure \ref{fig:cls} with Figure \ref{fig:isw} one can also notice that for a larger value of $\neff$ the early ISW causes an increase of power on the first and the second peaks, while the same variation in $\neff$ turns out in a reduction of power in the peaks at higher multipoles. This vertical shift is related to the Silk damping effect (dissusion damping in the baryon-photon plasma).

The decoupling of baryon-photon interactions is not instantaneous, but rather an extended process. This leads to diffusion damping of oscillations in the plasma, an effect known as Silk damping.
If decoupling starts at $\tau_{\rm d}$ and ends at $\tau_{\rm ls}$, during $\Delta\tau$
the radiation free streams on scale $\lambda_{\rm d}=\left(\lambda\Delta\tau\right)^{1/2}$
where $\lambda$ is the photon mean free path
and $\lambda_{\rm d}$ is shorter than the thickness of the last scattering surface.
As a consequence temperature fluctuations on scales smaller than $\lambda_{D}$ are damped,
because on such scales photons can spread freely both from overdensities and from underdensities.
The damping factor is $\exp[-(2r_{\rm d}/\lambda_{\rm d})]$ where $r_{\rm d}$ is the mean square diffusion distance at recombination. An approximated expression of $r_{\rm d}$ is given by \cite{Hou:2011ec}:
\begin{displaymath}
r_{\rm d}^2 = (2\pi)^2 \int_0 ^{a_{\rm ls}}{\frac{da}{a^3 \sigma_T n_e H}\left[\frac{R^2+\frac{6}{15}(1+R)}{6(1+R^2)}\right]}
\end{displaymath}
where $\sigma_T$ is the Thompson cross section, $n_e$ is the number density of free electrons, $a_{\rm ls}$ is the scale factor at recombination and the factor in square brackets is related to polarization \cite{Zaldarriaga:1995gi}. This diffusion process becomes more and more effective as last scattering is approached, so we can consider $a_{\rm}$ constant and thus obtain $r_{\rm d}\propto1/\sqrt{H}$.
Recalling the dependence $r_{\rm s}\propto1/H$ and the fact that $\theta_{\rm s}=r_{\rm s}/D_{\rm A}$ is fixed by CMB observations, we can infer $D_{\rm A}\propto1/H$. The result is that the damping angular scale $\theta_{\rm d}=r_{\rm d}/D_{\rm A}$ is proportional to the square root of the expansion rate $\theta_{\rm d}\propto\sqrt{H}$ and consequently it increases with the number of relativistic species. The effect on the CMB power spectrum can be seen in Figure \ref{fig:cls} bottom panel, where, in addition to the $\ell$ rescaling, we have subtracted the ISW of Figure \ref{fig:isw}. This damping effect shows up as a suppression of the peaks and a smearing of the oscillations that intensifies at higher multipoles.

It is important to stress that all these effects on the redshift of equivalence, on the size of the sound horizon at recombination, and on the damping tail can be compensated by varying other cosmological parameters. For instance the damping scale is affected by the helium fraction as well as by the effective number of relativistic degrees of freedom: $r_{\rm d}\propto(1-\yhe)^{-0.5}$ \cite{Hou:2011ec}. Therefore, at the level of the damping in the power spectrum, a larger value of $\neff$ can be mimicked by a lower value of $\yhe$ (see Figure \ref{fig:neffyhe}, Section \ref{sec:neffyhe}). The redshift of the equivalence $z_{\rm eq}$ can be kept fixed by increasing the cold dark matter density while inscreasing $\neff$. Finally an open Universe with a non zero curvature can reproduce the same peak shifting of a larger number of relativistic degrees of freedom. All these degeneracies increase the uncertainty of the results and degrade the constraint on $\neff$.

The only effect that cannot be mimicked by other cosmological parameters is the neutrino anisotropic stress.
The anisotropic stress arises from the quadrupole moment of the cosmic neutrino background temperature distribution and it alters the gravitational potentials \cite{Bashinsky:2003tk,Hannestad:2004qu}.
The effect on the CMB power spectrum is located at scales that cross the horizon before the matter-radiation equivalence ($\ell>\sim130$) and it consists of an increase in power by a factor $5/(1+\frac{4}{15}f_\nu)$ \cite{Hu:1995en}, where $f_\nu$ is the fraction of radiation density contributed by free-streaming particles.

\subsection{Dark radiation models}
\label{sec:models}
A number of theoretical physics models could explain a contribution to the extra dark radiation component of the universe, i.e. to $\Delta \neff$.

 A particularly simple model, based on neutrino oscillation short baseline physics results, contains sterile neutrinos. Sterile neutrinos are right handed fermions which do not interact  via any of the fundamental standard model interactions and therefore their number is not determined by
any fundamental symmetry in nature. Originally, models with one additional massive mainly sterile neutrino $\nu_4$, with a mass splitting $\Delta m^2_{14}$, i.e. the so called (3+1) models, were introduced to explain LSND~(Large Scintillator Neutrino Detector) \cite{Aguilar:2001ty} short baseline (SBL) antineutrino data by means of neutrino oscillations~\cite{3plus1}. A much better fit to both appearance and disappearance data was in principle provided by the (3+2) models~\cite{Sorel:2003hf} in which there are two mostly sterile neutrino mass states $\nu_4$ and $\nu_5$ with mass splittings in the range $0.1$~eV$^2< |\Delta m^2_{14}|, |\Delta m^2_{15}|< 10$~eV$^2$.  In the two sterile neutrino scenario we can distinguish two possibilities, one in which both mass splittings are positive, named as 3+2, and one in which one of them is negative, named as 1+3+1~\cite{Goswami:2007kv}.  Recent MiniBooNE antineutrino data are consistent with oscillations in the
$0.1$~eV$^2< |\Delta m^2_{14}|, |\Delta m^2_{15}| < 10$~eV$^2$, showing some overlapping with LSND results~\cite{Aguilar-Arevalo:2013pmq}. The running in the neutrino mode also shows an excess at low energy. However, the former excess seems to be not compatible with a simple two neutrino oscillation formalism~\cite{Aguilar-Arevalo:2013pmq}. A recent global fit to long baseline, short baseline, solar, and atmospheric neutrino oscillation data~\cite{Kopp:2013vaa} has shown that in the $3+1$ and $3+2$ sterile neutrino schemes there is some tension in the combined fit to appearance and disappearance data. This tension is alleviated in the $1+3+1$ sterile neutrino model case with a $p$ value of $0.2\%$. These results are in good agreement with those presented in Ref.~\cite{Conrad:2012qt}, which also considered the $3+3$ sterile neutrino models with three active and three sterile neutrinos. They conclude that $3+3$ neutrino models yield a compatibility of $90\%$ among all short baseline data sets-highly superior to those obtained in models with either one or two sterile neutrino species. The existence of this extra sterile neutrinos states can be in tension with BBN (see Sec.~\ref{sec:yhe}). However, the extra neutrino species may not necessarily be fully thermalised in the early universe. Even though the masses and mixing angles necessary to explain oscillation data would seem to indicate full thermalisation, the presence of e.g.\ a lepton asymmetry can block sterile neutrinop production and lead to a significantly lower final abundance, making the model compatible with BBN bounds, see Refs.~\cite{Dodelson:2005tp,Melchiorri:2008gq,leptas,cosmosbl,Mirizzi:2013kva}.

However, an extra radiation component may arise from many other physical mechanisms, as, for instance, QCD thermal axions or extended dark sectors with additional relativistic degrees of freedom. Both possibilities are closely related to minimal extensions to the standard model of elementary particles. Cosmological data provide a unique opportunity to place limits on any model containing new light species, see Ref.~\cite{Brust:2013ova}

We first briefly review the hadronic axion model~\cite{Kim:1979if,Shifman:1979if} since these hypothetical particles provide the most elegant and promising solution to the strong CP problem. Quantum Chromodynamics (QCD) respects CP symmetry, despite the existence of a natural, four dimensional, Lorentz and gauge invariant operator which violates CP. The presence of this CP violating-term will induce a non-vanishing neutron dipole moment, $d_n$. However, the experimental bound on the dipole moment $|d_{n}| < 3 \times 10^{-26}\ e \,$cm~\cite{dipole} would require a negligible CP violation contribution. Peccei and Quinn~\cite{PecceiQuinn} introduced a new
global $U(1)_{PQ}$ symmetry, which is spontaneously broken at a scale $f_a$, generating a new spinless particle, the axion.
The axion mass is inversely proportional to the axion decay constant $f_a$ which is the parameter controlling the interaction strength with the standard model plasma and therefore the degree of thermalisation in the early universe. The interaction Lagrangian is proportional to $1/f_a$ and high mass axions therefore have a stronger coupling to the standard model and thermalise more easily. Axions produced via thermal processes in the early
constitute providing a possible (sub)dominant  hot dark matter candidate, similar, but not exactly equivalent to, neutrino hot dark matter. High mass axions are disfavoured by cosmological data, with the specific numbers depending on the model and data sets used (see e.g. \cite{Hannestad:2005df,Hannestad:2007dd,Melchiorri:2007cd,DiValentino:2013qma}). Even though moderate mass axions can still provide a contribution to the energy density we also stress that just as for neutrino hot dark matter it cannot be mapped exactly to a change in $\neff$.

Generally, any model with a dark sector with relativistic degrees of freedom that eventually decouple from the standard model sector will also
contribute to $\neff$. Examples are the
asymmetric dark matter scenarios (see e.g. Refs.~\cite{Blennow:2010qp,Blennow:2012de} and
references therein), or extended weakly-interacting massive
particle models (see the recent work presented in Ref.~ \cite{Franca:2013zxa}).
We will review here the expressions from Ref. ~\cite{Blennow:2012de},  in which the authors include
 both light ($g_\ell$) and heavy ($g_h$)  relativistic degrees of freedom at the temperature of decoupling $T_D$  from the standard
 model.  For high decoupling temperature,  $T_D >$ MeV, the dark sector contribution to $\neff$ reads~\cite{Blennow:2012de}
\begin{displaymath}
\Delta \neff =\frac{13.56}{g_{\star S} (T_D)^\frac{4}{3}} \frac{(g_\ell+ g_h )^\frac{4}{3}}{g_\ell^\frac{1}{3}}~,
\end{displaymath}
where $g_{\star S} (T_D)$ refers to the effective number of entropy degrees of freedom at the dark sector decoupling temperature.
If the dark sector decouples at lower temperatures ($T_D <$ MeV),
there are two possibilities for the couplings of the dark sector
with the standard model: either the dark sector couples to the
electromagnetic plasma or it couples to neutrinos. In this former case,
\begin{displaymath}
\neff=\left(3+\frac{4}{7}\frac{(g_h+g_\ell)^\frac{4}{3}}{g_\ell^\frac{1}{3}}\right)\left(\frac{3\times \frac{7}{4}+g_H+g_h+g\ell}{3 \times \frac{7}{4}+g_h+g_\ell}\right)^\frac{4}{3}~,
\end{displaymath}
being $g_H$ the number of degrees of freedom  that become non relativistic between typical BBN temperatures and $T_D$.
The authors of \cite{Blennow:2012de} have shown that the cosmological constraints on $\neff$ can be translated into the required heavy degrees
of freedom heating the light dark sector plasma $g_h$ as a function of the dark sector decoupling temperature $T_D$
for a fixed value of $g_\ell$. Recent Planck data~\cite{Ade:2013zuv}, combined with measurements of the Hubble constant $H_0$
from the Hubble Space Telescope (HST), low multipole polarization measurements from the Wilkinson Microwave
Anisotropy Probe (WMAP) 9 year data release~\cite{Hinshaw:2012fq}  and high multipole CMB data from both the Atacama Cosmology Telescope
(ACT)~\cite{Sievers:2013ica} and the South Pole Telescope (SPT)~\cite{Hou:2012xq,Story:2012wx} provide the constraint $\neff$ is $3.62^{+0.50}_{-0.48}$.
Using this constraint, the authors of Ref.~\cite{DiValentino:2013qma} have found that having extra heavy degrees of freedom in the dark sector for low decoupling temperatures is highly disfavored.

Another aspect of dark radiation is that it could interact with the dark matter sector. In asymmetric dark matter models (see Ref.~\cite{Blennow:2010qp}), the dark matter production mechanism resembles to the one in the baryonic sector, with a particle-antiparticle asymmetry at high temperatures. The thermally symmetric dark matter component eventually annihilates and decays into dark radiation species. Due to the presence of such an interaction among the dark matter and dark radiation sectors, they behave as a tightly coupled fluid with pressure which will imprint oscillations in the matter power spectrum (as the acoustic oscillations in the photon-baryon fluid before the recombination era). The clustering properties of the dark radiation component may be modified within interacting schemes, and therefore the clustering parameters $\ceff$ and $\cvis$ may differ from their standard values for the neutrino case $\ceff = \cvis = 1/3$ (see Section \ref{sec:ceffcvis}).
In the presence of a dark radiation-dark matter interaction, the complete Euler equation for dark radiation, including the interaction term with dark matter, reads:
\begin{displaymath}
\dot{\theta}_{dr}=3k^2\ceff\left(\frac{1}{4}\delta_{dr}-\frac{\dot{a}}{a}\frac{\theta_{dr}}{k^2}\right)-\frac{\dot{a}}{a}\theta_{dr}-\frac{1}{2}k^2\pi_{dr}+an_{dm}\sigma_{dm-dr}(\theta_{dm}-\theta_{dr})~,
\end{displaymath}
where the term $an_{dm}\sigma_{dm-dr}(\theta_{dm}-\theta_{dr})$ represents the moment transferred to the dark radiation component and the quantity $an_{dm}\sigma_{dm-dr}$ gives the scattering rate of dark radiation by dark matter. The authors of \cite{Mangano:2006mp} have parametrized the coupling between dark radiation and dark matter through a cross section given by:
\begin{displaymath}
 \langle\sigma_{dm-dr}|v|\rangle \sim Q_0\, m_{dm}~,
\end{displaymath}
\noindent if it is constant, or
\begin{displaymath}
 \langle\sigma_{dm-dr}|v|\rangle \sim \frac{Q_2}{a^2}\, m_{dm}~,
\end{displaymath}
\noindent if it is proportional to $T^2$, where the parameters $Q_0$ and $Q_2$ are constants in cm$^2$ MeV$^{-1}$ units. It has been shown in Ref.~\cite{Blennow:2012de} that the cosmological implications of both constant and T-dependent interacting cross sections are very similar. Recent cosmological constraints on generalized interacting dark radiation models have been presented in Ref.~\cite{diamanti}.

{\bf got to here}

\section{Analysis Method}
\label{sec:method}

The parameter space (see Section \ref{sec:parameters}) is sampled through a Monte Carlo Markov Chain performed with the  publicly available package \texttt{CosmoMC} \cite{Lewis:2002ah} based on the Metropolis-Hastings sampling algorithm and on the Gelman Rubin convergence diagnostic.
The calculation of the theoretical observables is done through \texttt{CAMB} \cite{Lewis:1999bs} (Code for Anisotropies in the Microwave Background) software. The code is able to fit any kind of cosmological data
with a bayesian statistic,
in our case we focus on the data sets reported in the following Section.

\subsection{Data sets}
\label{sec:datasets}

Our basic data set is the Planck temperature power spectrum (both at low $\ell$ and at high $\ell$) in combination with the WMAP 9 year polarization data (hereafter WP) and the high multipole CMB data of ACT and SPT (hereafter highL). These data sets are implemented in the analysis following the prescription of the Planck likelihood described in \cite{Planck:2013kta}.
The additional data sets test the robustness at low redshift of the predictions obtained with CMB data. These data sets consist of a prior on the Hubble constant from the Hubble Space Telescope measurements \cite{Riess:2011yx} (hereafter $H_0$) and the information on the dark matter clustering from the matter power spectrum
extracted from the Data Release 9 (DR9)  of the CMASS sample of
galaxies \cite{Ahn:2012fh} from the Baryon Acoustic Spectroscopic
Survey (BOSS)\cite{Dawson:2012va} part of the program of the Sloan
Digital Sky Survey III~\cite{Eisenstein:2011sa}.

\subsection{Parameters}
\label{sec:parameters}

In Table \ref{tab:priors} the parameters used in the analyses are listed together with the top-hat priors on them.
The six standard parameters of the $\Lambda$CDM model are:
the physical baryon density, $\omega_b\equiv\Omega_bh^{2}$;
the physical cold dark matter density, $\omega_c\equiv\Omega_ch^{2}$;
the angular scale of the sound horizon, $\theta_{\rm s}$;
the reionization optical depth, $\tau$;
the amplitude of the primordial spectrum at a certain pivot scale, $A_{s}$;
the power law spectral index of primordial density (scalar) perturbations, $n_s$.

We include the effective number of relativistic degrees of freedom $\neff$, and,
in addition, our runs also contain one, or a combination, of the following parameters:
the sum of neutrino masses $\summnu$, the primordial helium fraction $\yhe$, the neutrino perturbation parameters, namely the effective sound speed $\ceff$ and the viscosity parameter $\cvis$. Finally we also investigated the impact of a varying lensing amplitude $A_L$.

We assume that massive neutrinos are degenerate and share the same mass. Indeed given the present accuracy of CMB measurements, cosmology cannot extract the neutrino mass hierarchy, but only the total hot dark matter density. However the future measurements of the Euclid survey will achieve an extreme accurate measurement of the neutrino mass ($\sigma m_\nu \simeq 0.01$~eV \cite{Basse:2013zua}), which will pin down the neutrino mass hierarchy.

\subsubsection{Primordial helium fraction}
\label{sec:yhe}
The primordial helium fraction, $\yhe$, is a probe of the number of relativistic species at the time of Big Bang Nucleosynthesis. As we have seen in Section \ref{sec:effects}, when $\neff$ increases the expansion rate increases as well. This means that free neutrons have less time to convert to protons through $\beta$ decay before the freeze out and so the final neutron to proton ratio is larger. The observable consequence is that the helium fraction is higher.

Measurements of the primordial light element abundances seem consistent with a standard number of relativistic species at the time of BBN at 95\% c.l. ($\neff^{\rm BBN}=3.68^{+0.80}_{-0.70}$ \cite{Izotov:2010ca}).
However the value of $\neff$ at BBN ($T\sim1$~MeV) and the value measured by CMB at the last scattering epoch ($T\sim1$eV) may be different because of the unknown physics in the region 1~Mev$<T<$1~eV. Several efforts have been carried out in order to solve the tension among $\neff^{\rm BBN}$ and $\neff^{\rm CMB}$:
decay of massive particles ($1$~MeV$<m<1$~eV) in additional relativistic species \cite{Fischler:2010xz,Bjaelde:2012wi},
decay of gravitino into axino and axion \cite{Hasenkamp:2011em},
or neutrino asymmetries \cite{Krauss:2010xg}.

The BBN consistency relation implies that the number of relativistic species present at BBN is the same as the number measured by CMB at recombination. In order to impose the BBN consistency we use the standard option implemented in \texttt{CosmoMC} \cite{Hamann:2007sb}. This routine calculates $\yhe$ as a function of $\neff$ and $\Omega_b h^2$ using a fitting formula obtained with the \texttt{ParthENoPE} code \cite{Pisanti:2007hk}.

\subsubsection{Lensing amplitude}
\label{sec:alens}

Massive neutrinos suppress the growth of dark matter perturbations both through free streaming and through the equivalence delay. As a consequence the matter power spectrum is damped on scales smaller than the scale of the horizon when neutrino become non relativistic. The accuracy level of Planck allows for a detection of this clustering suppression in the CMB lensing potential, so it is timely to investigate the correlation among $A_L$ and neutrino parameters.
Planck analysis \cite{Ade:2013zuv} provides an anomalous value of the lensing amplitude $A_L=1.23\pm0.11$ (68\% c.l., Planck+WP+highL). This anomaly was already revealed by ACT data ($A_L=1.70\pm0.38$ \cite{Sievers:2013ica}), but it is in tension with the SPT value $A_L=0.86^{+0.15}_{-0.13}$ \cite{Story:2012wx}, which is consistent with the standard prediction $A_L=1$. Subsequent analyses \cite{Said:2013hta} have confirmed this anomaly and studied the impact on massless $\neff$.

Even if a modification of General Relativity cannot be ruled-out, this anomaly is most likely a spurious signal related to the bias induced by the combination of data sets belonging to different experiments with different experimental techniques and different analysis methods. However it is important to account for its effect in order to get unbiased constraints on the sum of neutrino masses, that, as we will see in Section \ref{sec:neffmnu}, is the most correlated parameter with $A_L$.

\subsubsection{Neutrino perturbation parameters}
\label{sec:ceffcvis}

As we have seen in Section \ref{sec:models}, there is a wide variety of models that can explain an excess in the number of relativistic degrees of freedom at decoupling.
In order to distinguish between these models, we introduce the neutrino perturbation parameters, the effective sound speed and the viscosity parameter, $\ceff$ and $\cvis$, respectively \cite{Hu:1998kj}.
The reason is that these parameters can characterize the properties of the component that accounts for extra relativistic species.

Following \cite{Archidiacono:2011gq} and \cite{Smith:2011es}, we encode $\ceff$ and $\cvis$ in the massless neutrino perturbation equations:
\begin{align*}
&\dot{\delta}_{\nu} =  \frac{\dot{a}}{a} (1-3 \ceff) \left(\delta_{\nu}+3 \frac{\dot{a}}{a}\frac{q_{\nu}}{k}\right)-k \left(q_{\nu}+\frac{2}{3k} \dot{h}\right), \\
&\dot{q}_{\nu}  =  k \ceff \left(\delta_{\nu}+3 \frac{\dot{a}}{a}\frac{q_{\nu}}{k}\right)- \frac{\dot{a}}{a}q_{\nu}- \frac{2}{3} k \pi_{\nu}, \\
&\dot{\pi}_{\nu} =  3 \cvis \left(\frac{2}{5} q_{\nu} + \frac{8}{15} \sigma\right)-\frac{3}{5} k F_{\nu,3}, \\
&\frac{2l+1}{k}\dot{F}_{\nu,l} -l F_{\nu,l-1} =  - (l+1) F_{\nu,l+1},\ l \geq 3 \ .
\end{align*}
Here the equations are written in the synchronous gauge (the one used in \texttt{CAMB} package \cite{Lewis:1999bs}), the dot indicates the derivative respect to conformal time $\tau$, $a$ is the scale factor, $k$ is the wavenumber, $\delta_{\nu}$ is the neutrino density contrast, $q_{\nu}$ is the neutrino velocity perturbation, $\pi_{\nu}$ is the neutrino anisotropic stress, $F_{\nu, \ell}$ are higher order moments of the neutrino distribution function and $\sigma$ is the shear.

The viscosity parameter is related to the clustering properties of particles, because it parametrizes the relationship between velocity/metric shear and anisotropic stress: $\cvis=0$ indicates a perfect fluid with undamped perturbations, while an increased value of $\cvis$ makes the oscillations related to this component to be overdamped. Free streaming particles, such as neutrinos, lead to anisotropies in the Cosmic Neutrino Background that are characterized by $\cvis=1/3$.

When $\ceff$ decreases the internal pressure of the dark radiation fluid decreases and the its perturbations can grow and cluster; on the contrary, if $\ceff$ increases the oscillations are damped.
Furthermore an increase (decrease) in $\ceff$ leads to an increase (decrease) in the neutrino sound horizon and, as a consequence, also in the scale at which neutrino perturbations affect the dark radiation fluid.

If the additional relativistic species we are dealing with consist of free streaming particles, such as neutrinos, the neutrino perturbation parameters would be $\ceff=\cvis=1/3$.

\begin{table}[t]
\begin{center}
\begin{tabular}{lc}
\hline
\hline
 Parameter & Prior\\
\hline
$\omega_{\rm b}$ & $0.005 \to 0.1$\\
$\omega_{\rm cdm}$ & $0.001 \to 0.99$\\
$\theta_{\rm s}$ & $0.5 \to 10$\\
$\tau$ & $0.01 \to 0.8$\\
$\ln{(10^{10} A_{\rm s})}$ & $2.7 \to 4$\\
$n_{\rm s}$ & $0.9 \to 1.1$\\
\hline
$A_{L}$ & $0 \to 5$\\
$\neff$ & $0 \to 7$\\
$\summnu$ [eV] &  $0 \to 7$\\
$\yhe$ & $0.1 \to 0.5$ \\
$\ceff$ & $0 \to 1$\\
$\cvis$ & $0 \to 1$\\
\hline
\hline
\end{tabular}
\caption{Priors for the cosmological parameters considered for the fits in this work. All priors are uniform (top hat) in the given intervals.}
\label{tab:priors}
\end{center}
\end{table}

\section{Results}
\label{sec:results}

In what follows the results of our analyses are presented. These results cover a wide range of different parameter spaces and they are obtained using different combinations of data sets. In Section \ref{sec:neffyhe} we study the impact of a varying Helium fraction and of the BBN consistency relation on the effective number of relativistic degrees of freedom. Section \ref{sec:neffmnu} analyses the dependence of the neutrino abundances and masses on the varying lensing amplitude and on the matter power spectrum information. Finally in Section \ref{sec:neffceffcvis} we provide constraints on the neutrino perturbation parameters.

\subsection{Constraints on $\neff$: number of relativistic species}
\label{sec:neffyhe}

In Table \ref{tab:neffyhe} the constraints on the number of effective relativistic degrees of freedom are shown with different priors.
\begin{savenotes}
\begin{table*}[!h]
\begin{center}
\resizebox{1\textwidth}{!}{
\begin{tabular}{lccccc}
\hline
\hline
   &             Planck+WP+highL & Planck+WP+highL & Planck+WP+highL & Planck+WP+highL & Planck+WP+highL\\
   &                                          & + $H_0$                   & + BBNc                  & + $H_0$ + BBNc & \\
\hline
\hspace{1mm}\\
${\neff}$   & $3.63\pm0.41$ & $3.81\pm0.29$ & $3.44\pm0.35$ & $3.65\pm0.26$ & $3.32\pm0.70$\\
\hline
\hspace{1mm}\\
$\yhe$      & $0.24$ & $0.24$ & BBN & BBN & $0.260\pm0.036$ \\
\hline
\hline
\end{tabular}
}
\caption{Marginalized 68\% constraints on $\neff$ in a
  standard cosmology with $\neff$ massless neutrinos. In the second and in the fourth columns we also apply a prior on the Hubble constant from the Hubble Space Telescope measurements. In the third and in the fourth columns we apply the BBN consistency relation.}
\label{tab:neffyhe}
\end{center}
\end{table*}
\end{savenotes}

\begin{figure*}[!h]
\begin{tabular}{ccc}
\includegraphics[width=5cm]{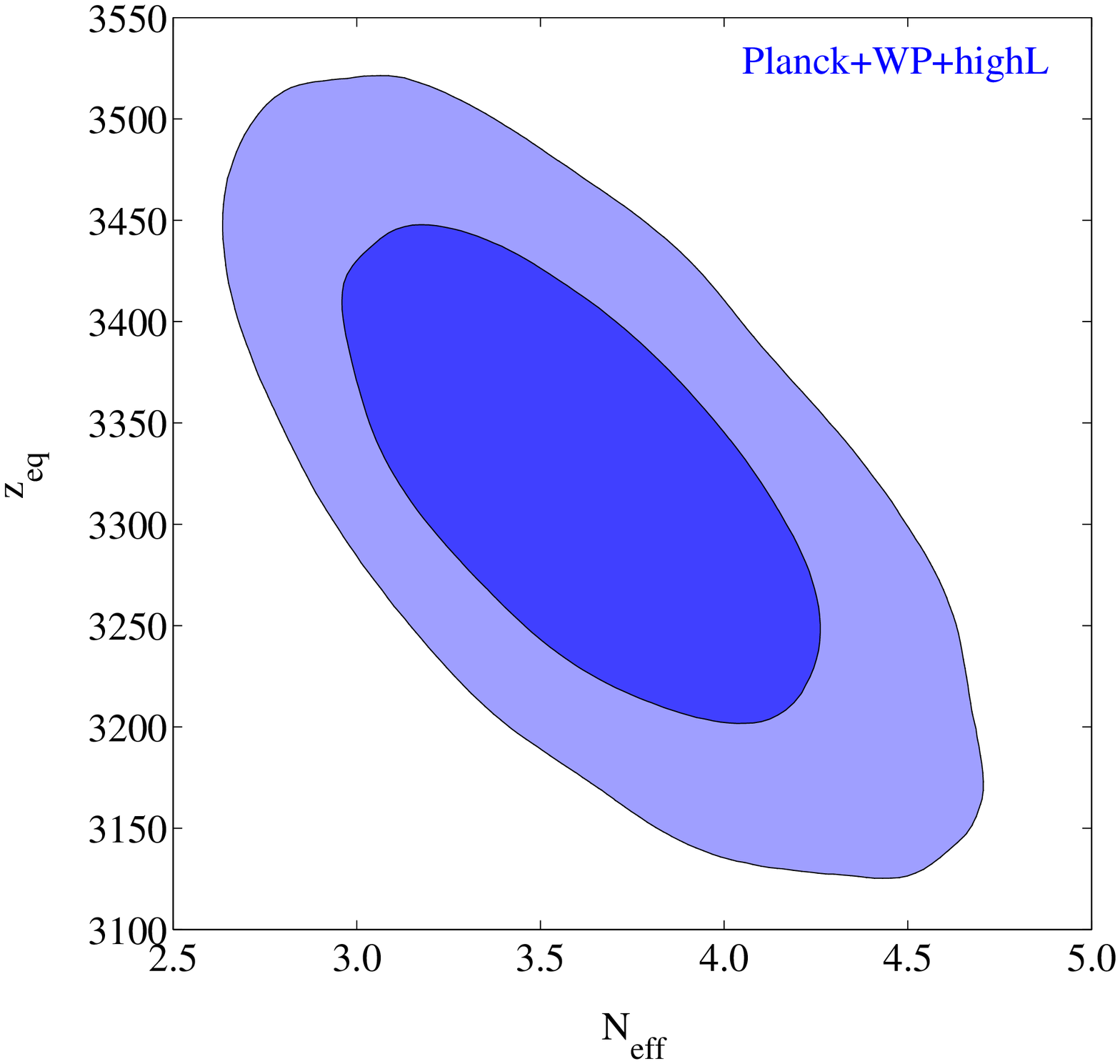}&\includegraphics[width=5cm]{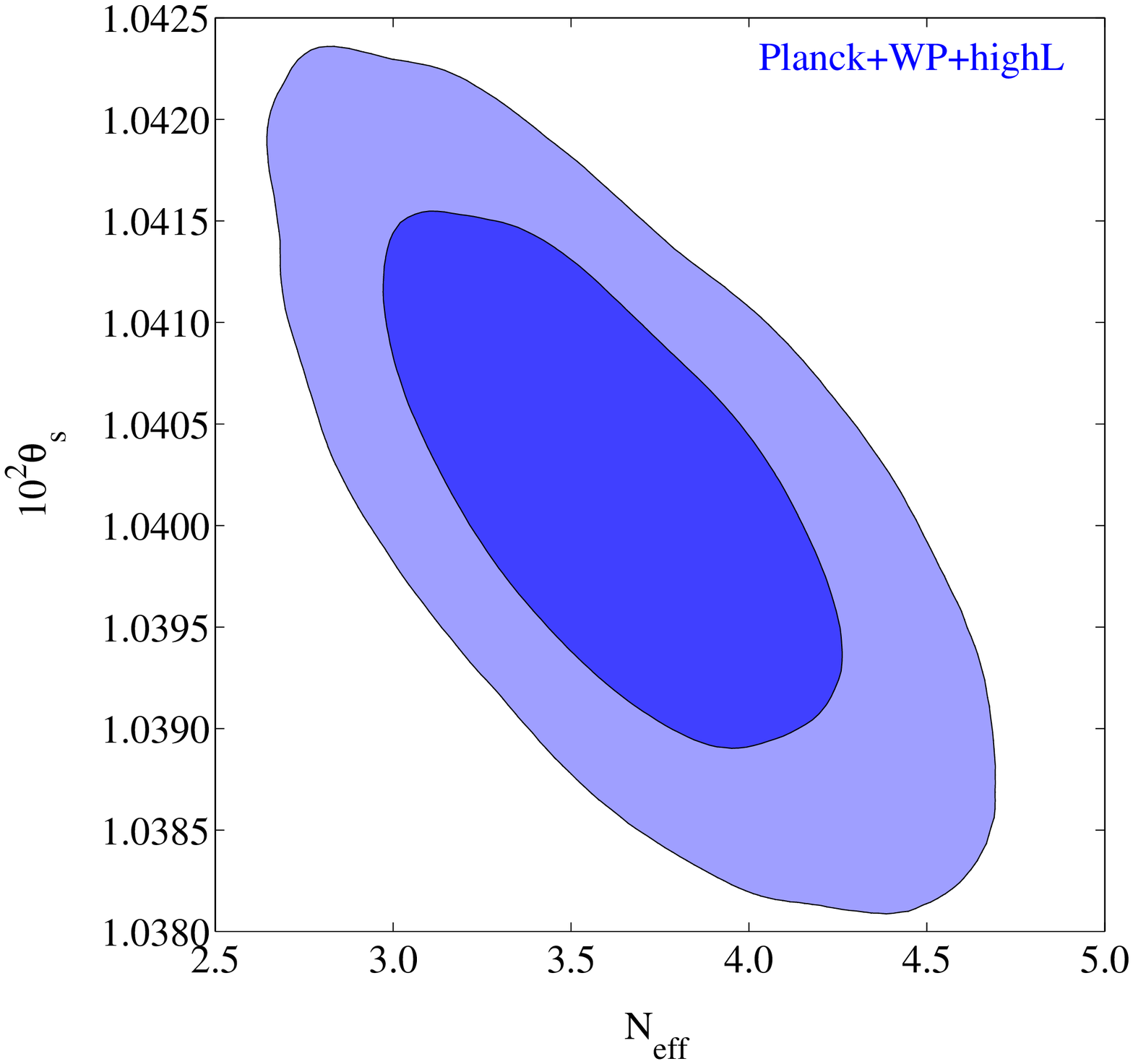}&\includegraphics[width=5cm]{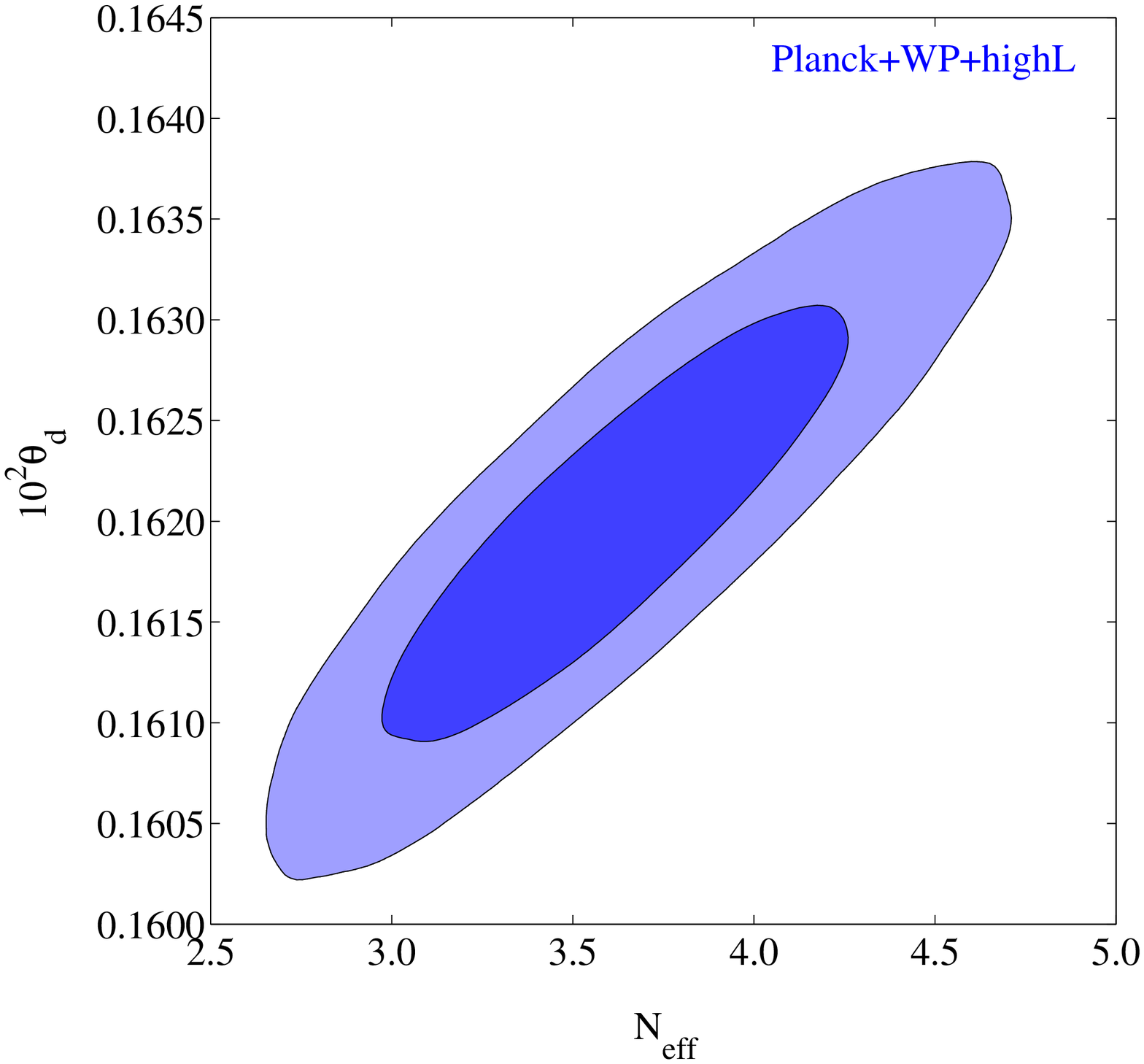}
\end{tabular}
 \caption{$68\%$ and $95\%$~c.l. 2D marginalized posterior in the plane $\neff - H_0$.}
\label{fig:neffeffects}
\end{figure*}

First of all, in order to recall the effects of the number of effective relativistic degrees of freedom on CMB, we show in Figures \ref{fig:neffeffects} the degeneracies among $\neff$ and the parameters that are directly measured by the CMB temperature power spectrum: the redshift of the equivalence $z_{\rm eq}$, the angular scale of the sound horizon $\theta_s$ and the damping scale $\theta_d$. We can notice that $z_{\rm eq}$ is proportional to the increase of $\neff$ as expected from equation \ref{eq:zeq}; while $\theta_d$ is correlated to $\neff$ through the expansion rate at recombination $H$, because it scales as $\sqrt{H}$.

\begin{figure*}[!h]
\begin{center}
\includegraphics[width=10cm]{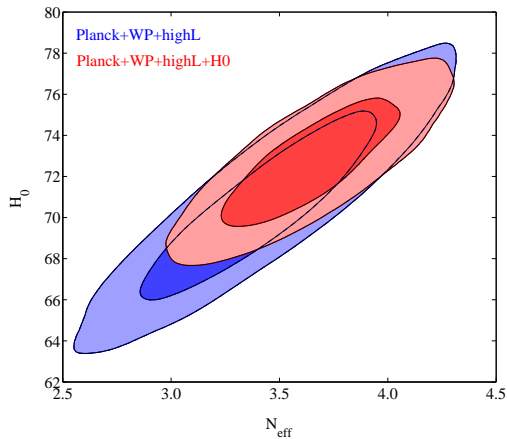}
\end{center}
 \caption{$68\%$ and $95\%$~c.l. 2D marginalized posterior in the plane $\neff - H_0$.}
\label{fig:neffh0}
\end{figure*}

The inclusion of the $H_0$ prior moves the mean of $\neff$ value toward a higher value and reduces the error on $\neff$ ($\neff=3.81\pm0.29$ respect to $\neff=3.63\pm0.41$, 68\% c.l.). The effect can be noticed in Figure \ref{fig:neffh0}. The final result is a $\sim2.6\sigma$ evidence for an extra dark radiation component.
Instead, applying the BBN consistency relation leads to a constraint on $\neff$ much closer to the standard value than in the case of $\yhe$ fixed to 0.24, i.e. $\neff=3.44\pm0.35$ (68\% c.l.).

\begin{figure*}[!h]
\begin{center}
\includegraphics[width=7cm]{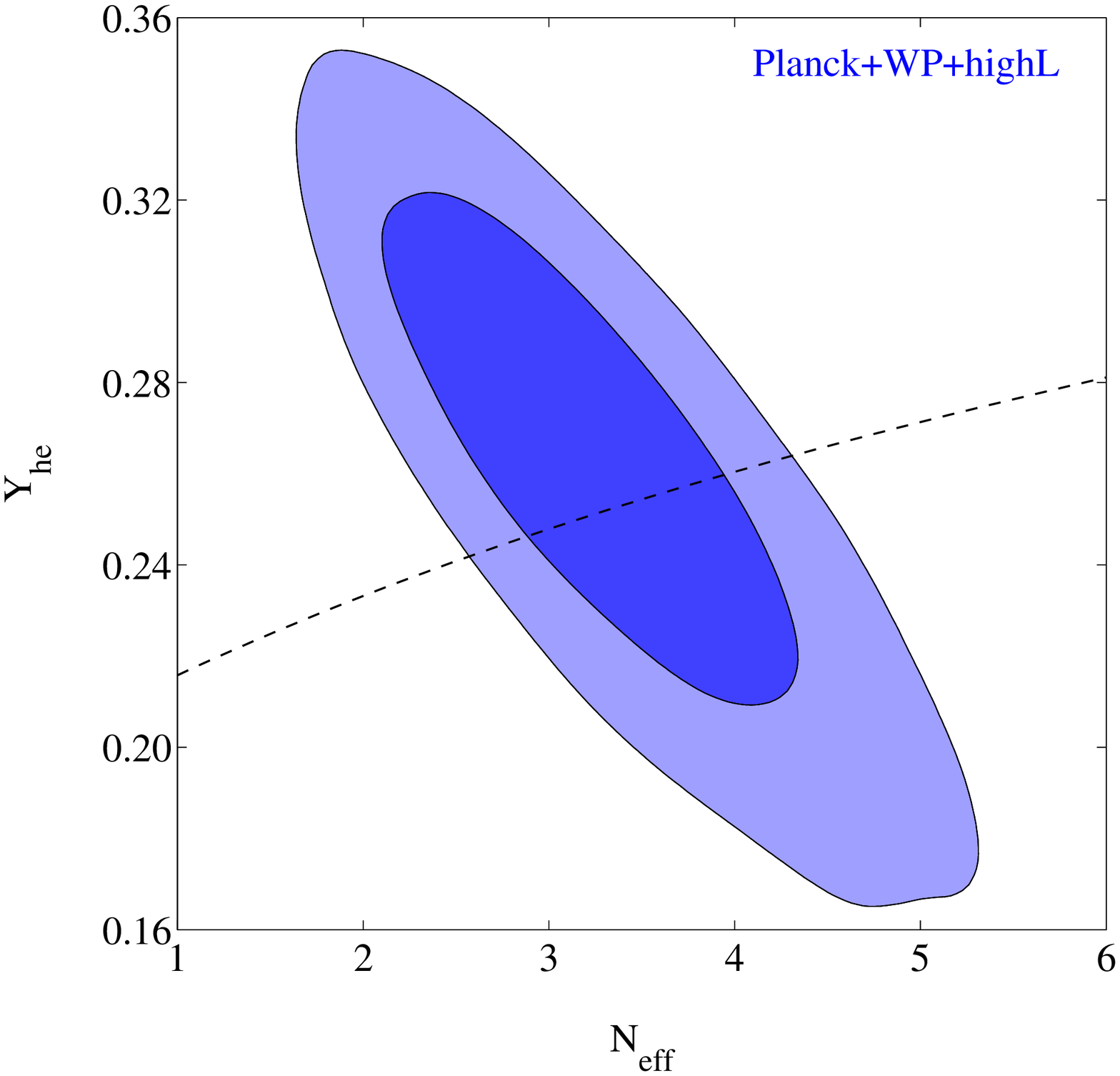}
 \caption{$68\%$ and $95\%$~c.l. 2D marginalized posterior in the plane $\neff - \yhe$.}
\label{fig:neffyhe}
\end{center}
\end{figure*}

Finally, if we consider the helium fraction as a free parameter (last column of Table \ref{tab:neffyhe}), the evidence for an extra number of relativistic degrees of freedom disappears and we obtain a milder constraint on $\neff$ ($\neff=3.32\pm0.70$, 68\% c.l.) that makes it perfectly consistent with the prediction of the Standard Model.
Figure \ref{fig:neffyhe} shows the anti correlation between $\neff$ and $\yhe$ from CMB data (blue contours) and the BBN consistency relation among these two parameters (dotted line). We can notice that an increase in $\neff$ requires a lower value of $\yhe$ to reproduce the same CMB power spectrum, as we have explained in Section \ref{sec:effects}.
Concerning the comparison between the models with and without varying the primordial helium fraction the $\Delta \chi^2$ at the best fit point is negligible, meaning that a higher value of $\yhe$ is preferred by the data but a lower value can be accommodated by tuning the other parameters.

\begin{figure*}[!h]
\begin{center}
\includegraphics[width=7cm]{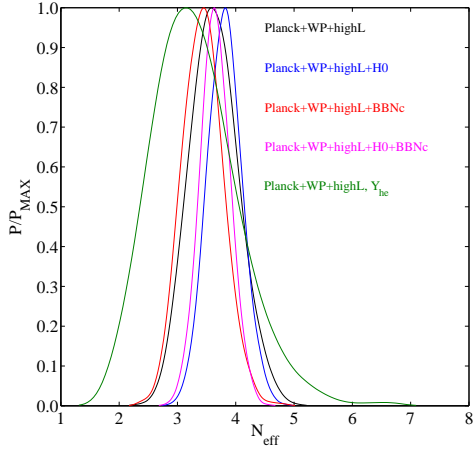}
 \caption{1D posterior of $\neff$. The different cases reported in Table \ref{tab:neffyhe} are shown: black line corresponds to Planck+WP+highL, blue line to Planck+WP+highL+$H_0$, red line to Planck+WP+highL+BBNc and magenta line to Planck+WP+highL+BBNc+$H_0$. Finally the green line refers to the analysis that includes also a varying $\yhe$.}
\label{fig:neff}
\end{center}
\end{figure*}
All the cases described above are illustrated in Figure \ref{fig:neff} where the one-dimensional posterior of $\neff$ is shown for the diferent cases of Table \ref{tab:neffyhe}. We can notice that both the inclusion of $H_0$ and BBN consistency narrow the posterior and reduce the error on $\neff$. However $H_0$ moves the best-fit of $\neff$ toward a higher value of the number of effective relativistic degrees of freedom, while BBN consistency prefers a lower value and brings back $\neff$ closer to the standard value.
In subsequent analyses we will follow a conservative approach, applying the BBN consistency relation in all our MCMC analyses, accordingly also with Planck team strategy.

\subsection{Constraints on $\neff$ and $\summnu$: massive neutrinos}
\label{sec:neffmnu}

The constraints on massive neutrinos are summarized in Table \ref{tab:neffmnualens}.

\begin{table*}[!h]
\begin{center}
\resizebox{1\textwidth}{!}{
\begin{tabular}{lccccc}
\hline
\hline
   & Planck+WP+highL & Planck+WP+highL & Planck+WP+highL &  Planck+WP+highL & Planck+WP+highL\\
    &                            &                               & +$H_0$                     &   +DR9                    & +DR9+$H_0$ \\
\hline
\hspace{1mm}\\
$\neff$   & $3.38\pm0.36$ & $3.65\pm0.38$ & $3.81\pm0.28$ & $3.33\pm0.31$ & $3.65\pm0.26$\\
\hline
\hspace{1mm}\\
$\summnu$ [eV]   & $<0.64$ & $<1.03$ & $<0.66$ & $<0.66$ & $<0.51$ \\
\hspace{1mm}\\
$A_{L}$   & $1$ & $1.36\pm0.14$ & $1.36\pm0.14$ & $1.10\pm0.08$ & $1.10\pm0.07$ \\
\hline
\hline
\end{tabular}
}
\caption{Marginalized 68\% constraints on $\neff$ and $A_L$
and 95\% cl upper bounds on $\summnu$
in extended models with $\neff$ massive neutrinos. We also include the lensing amplitude as a free parameter.
}
\label{tab:neffmnualens}
\end{center}
\end{table*}

\begin{figure*}[!t]
\begin{tabular}{cc}
\includegraphics[width=7cm]{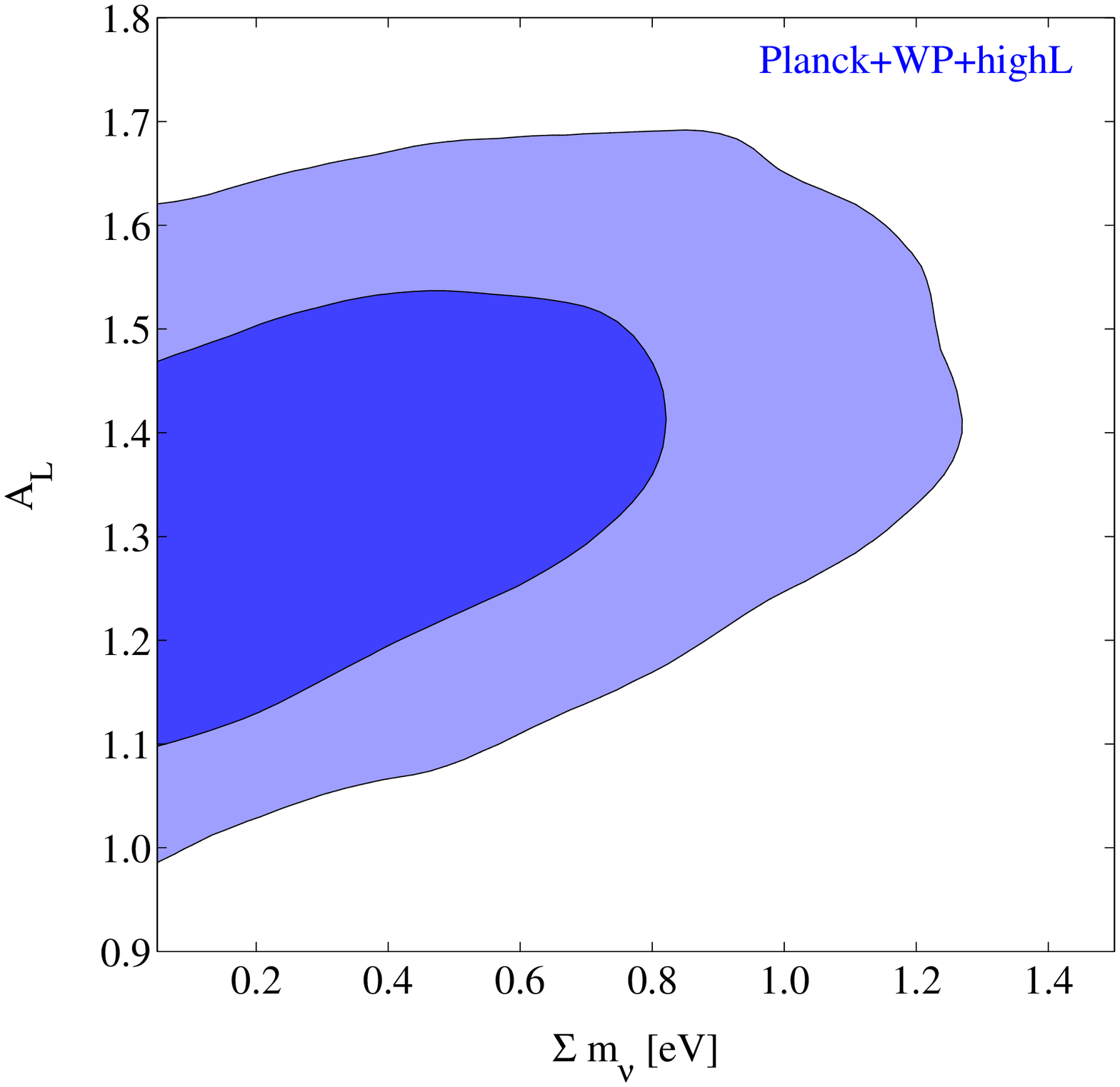} & \includegraphics[width=7cm]{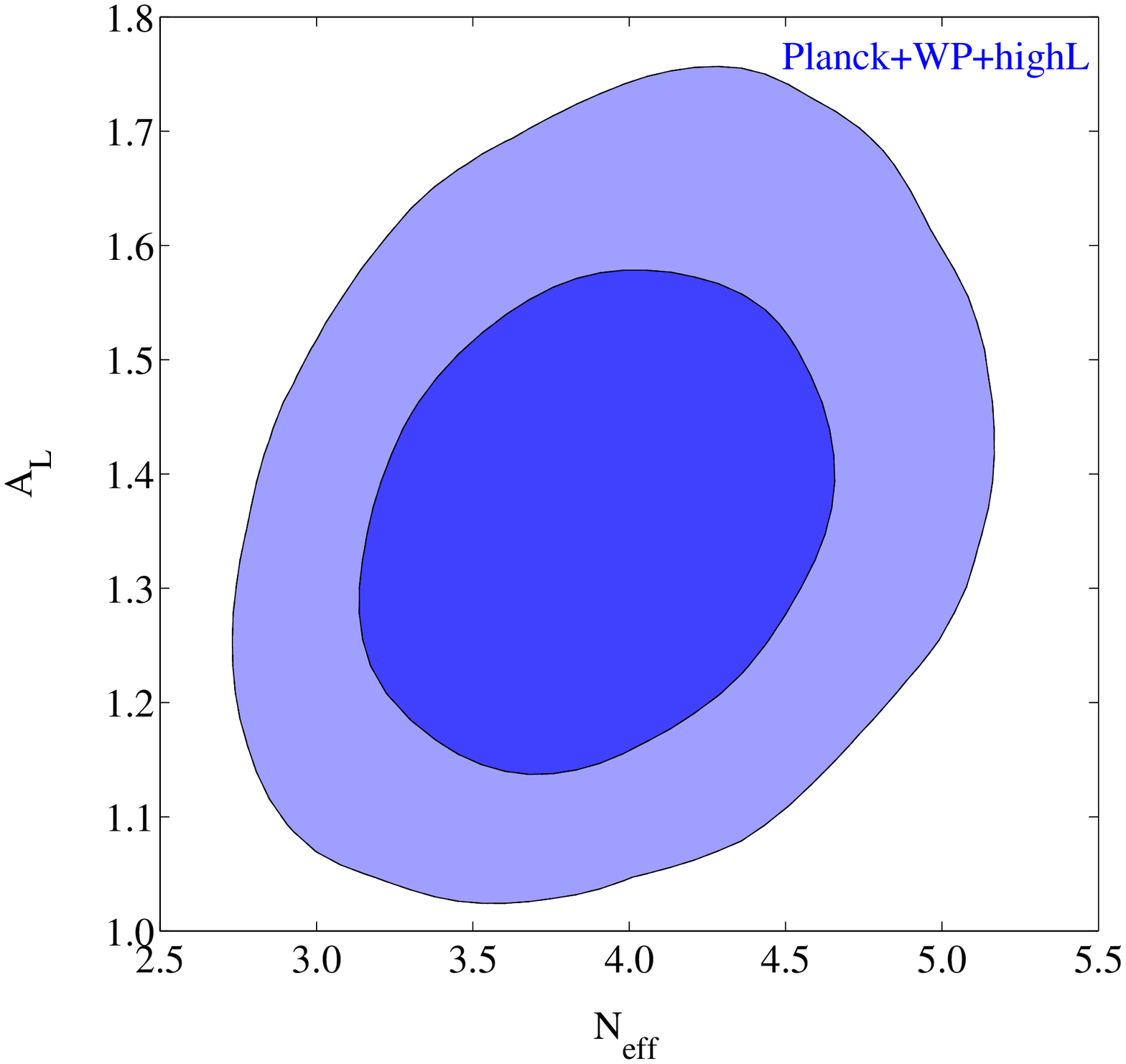}
\end{tabular}
 \caption{$68\%$ and $95\%$~c.l. 2D marginalized posterior in the plane $\summnu - A_{L}$ ({\it left panel}) and in the plane $\neff - A_{L}$ ({\it right panel}).}
\label{fig:alens}
\end{figure*}

\begin{figure*}[!h]
\begin{tabular}{cc}
\includegraphics[width=7cm]{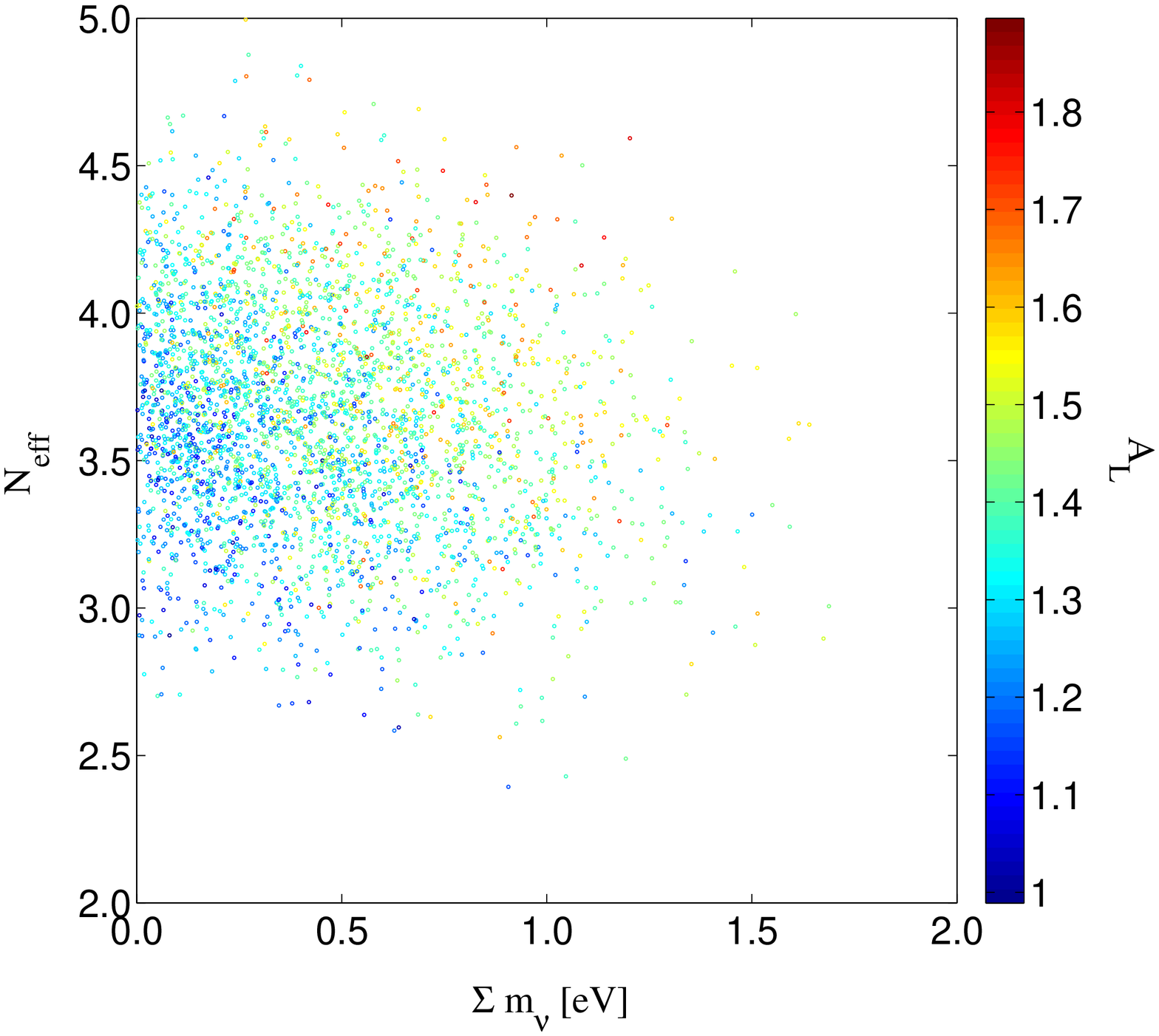} &\includegraphics[width=7cm]{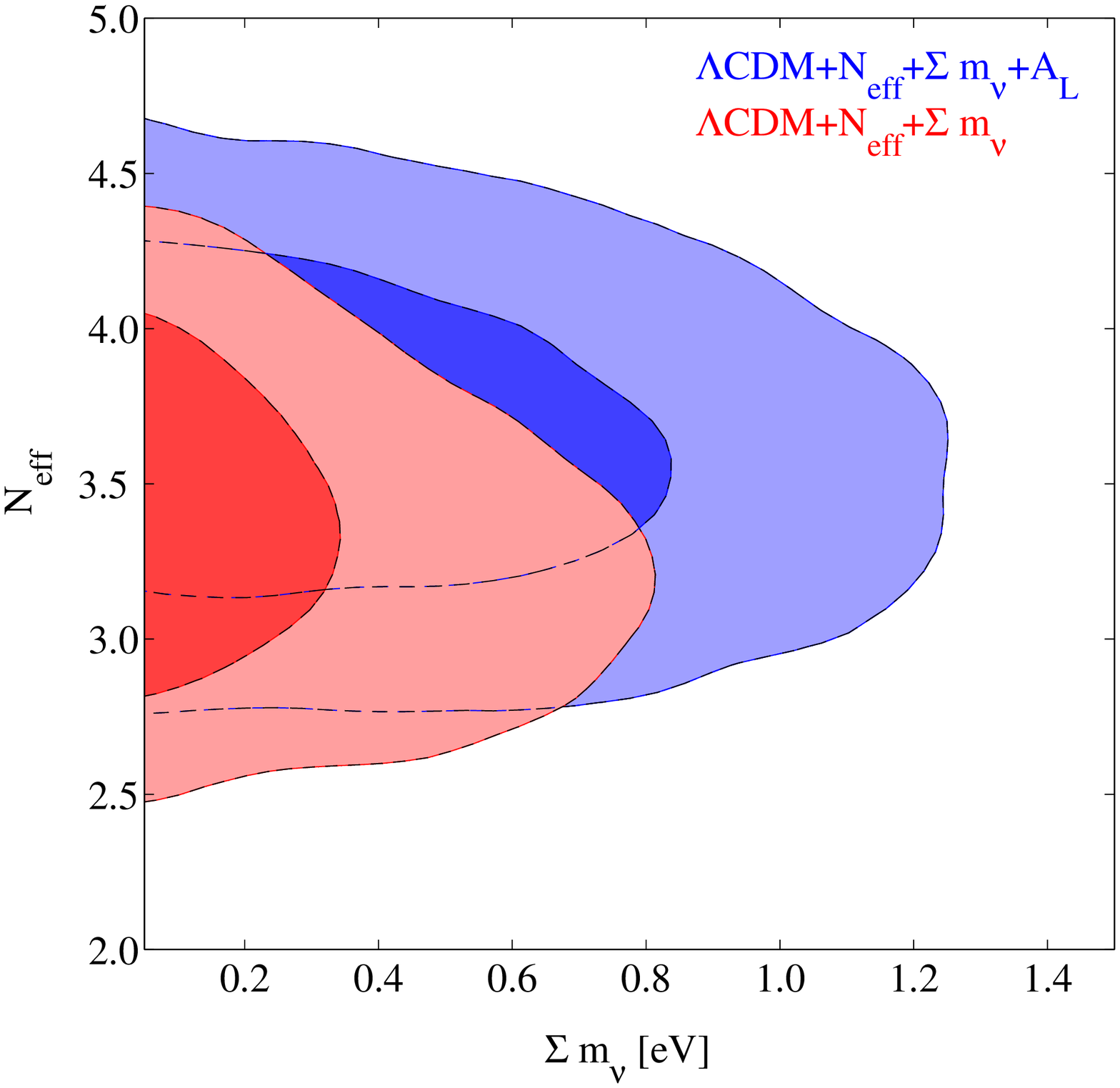}
\end{tabular}
 \caption{({\it left panel}) Scatter plot in the $\summnu - \neff$ plane with points colored by the value of the $A_L$ parameter (second column of Table \ref{tab:neffmnualens}). ({\it Right panel}) $68\%$ and $95\%$~c.l. 2D marginalized posterior in the plane $\summnu - \neff$, blue contours refer to the case with a varying lensing amplitude (second column of Table \ref{tab:neffmnualens}), red contours illustrate the $A_L=1$ case (first column of Table \ref{tab:neffmnualens}).}
\label{fig:neffmnualens}
\end{figure*}

We also marginalize over the lensing amplitude and we study this effect in Figure \ref{fig:alens} and in Figure \ref{fig:neffmnualens} for our basic data set (Planck+WP+highL).  As we already discussed in Section \ref{sec:alens} Planck analysis points towards a value of the lensing amplitude higher than the standard one. This anomaly is confirmed by our results $A_L=1.36\pm0.14$ (68\% c.l.) related to the model with a varying number of massive neutrinos. Nevertheless, including DR9 data shift the $A_L$ parameter towards a value consistent with the standard $A_L=1$ value within $2\sigma$ ($A_L=1.10\pm0.08$, 68\% c.l.). It is clear from the left panel of Figure \ref{fig:alens} that the neutrino mass has a strong degeneracy with the lensing amplitude: allowing for a higher value of $A_L$ leads to a larger value of the neutrino mass, the 95\% upper bound moves from $0.64$~eV to $1.03$~eV. The right panel of Figure \ref{fig:alens} shows that there is no preferred direction for a correlation between $\neff$ and $A_L$, but the side effect of the degeneracy among $\summnu$ and $A_L$ is also an increasing value of $\neff$ ($3.65\pm0.38$ against $3.38\pm0.36$, 68\% c.l.) related to the correlation among $\neff$ and $\summnu$. This conclusion arises from left panel of Figure \ref{fig:neffmnualens} where the increasing value of $A_L$ is located along the bisecting line in the plane $\neff - \summnu$. We summarize the effect of the lensing amplitude on the neutrino parameters in the right panel of Figure \ref{fig:neffmnualens}: a varying $A_L$ parameter will lead to a larger neutrino mass and, consequently, to a larger $\neff$. Finally, we shall comment that a larger value of $A_L$ will provide a better fit to the data, lowering the $\chi^2$ by 4.2 units.

\begin{figure*}[!h]
\begin{tabular}{ccc}
\includegraphics[width=5.cm]{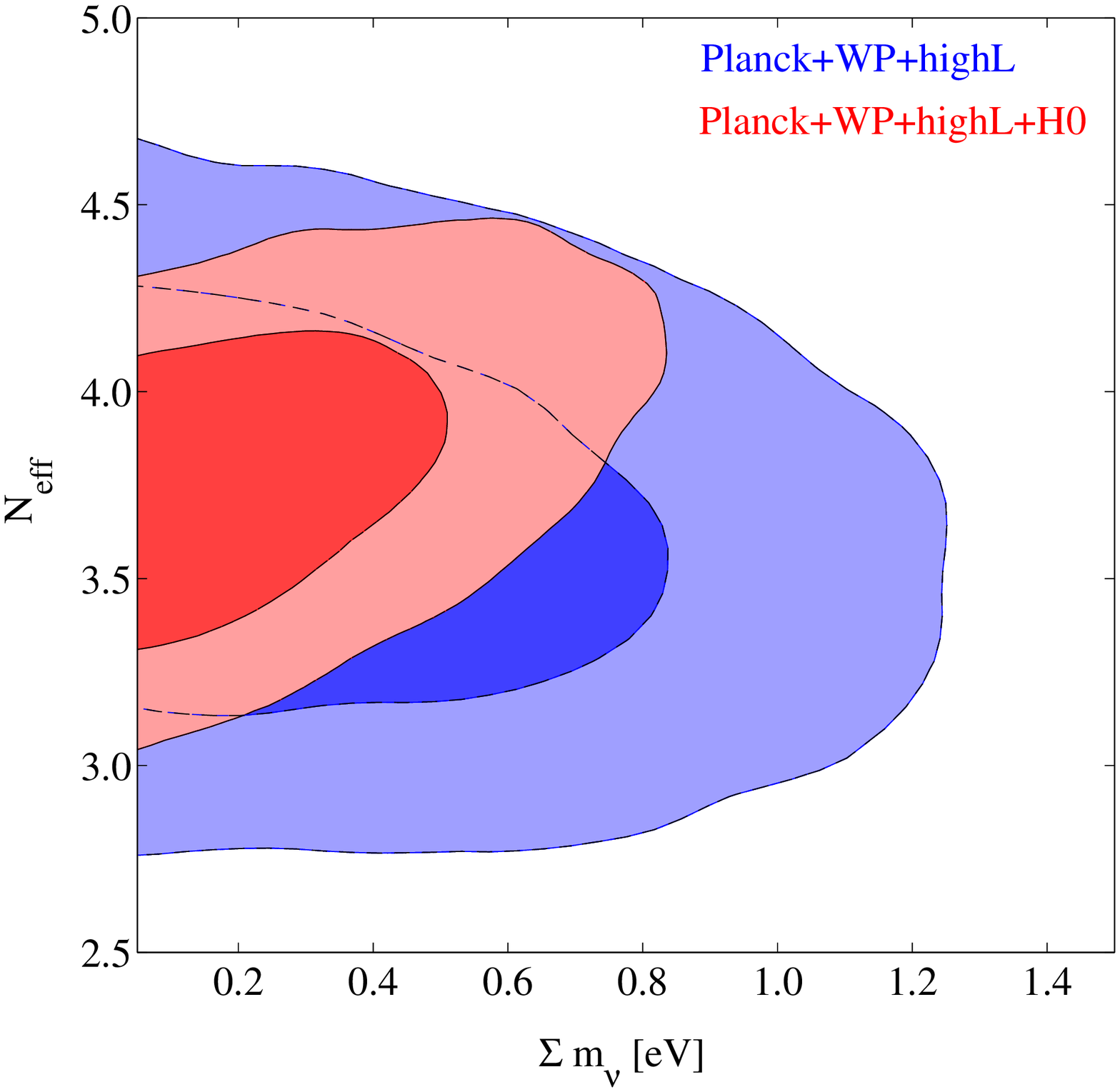} & \includegraphics[width=5.cm]{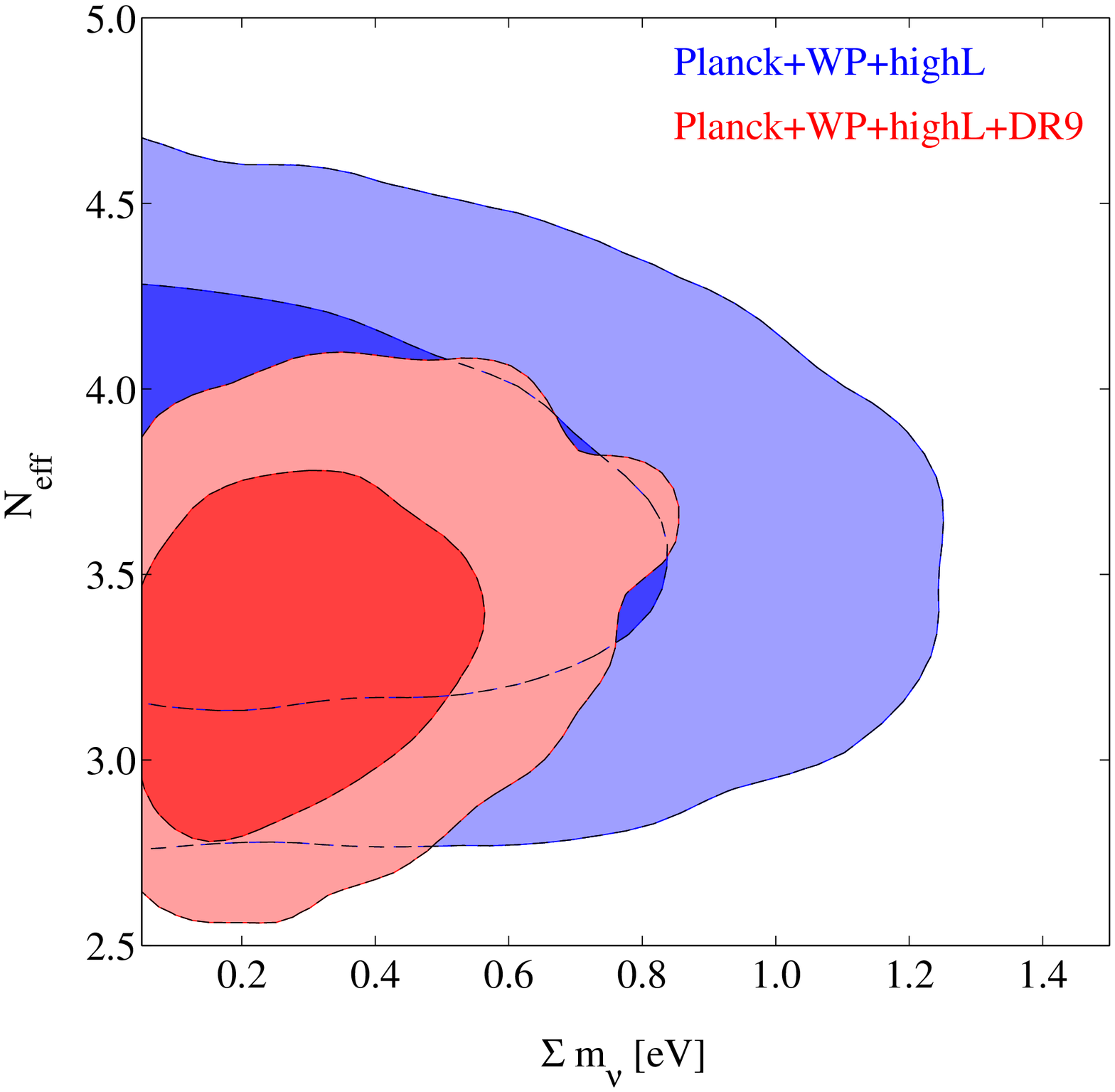} & \includegraphics[width=5.cm]{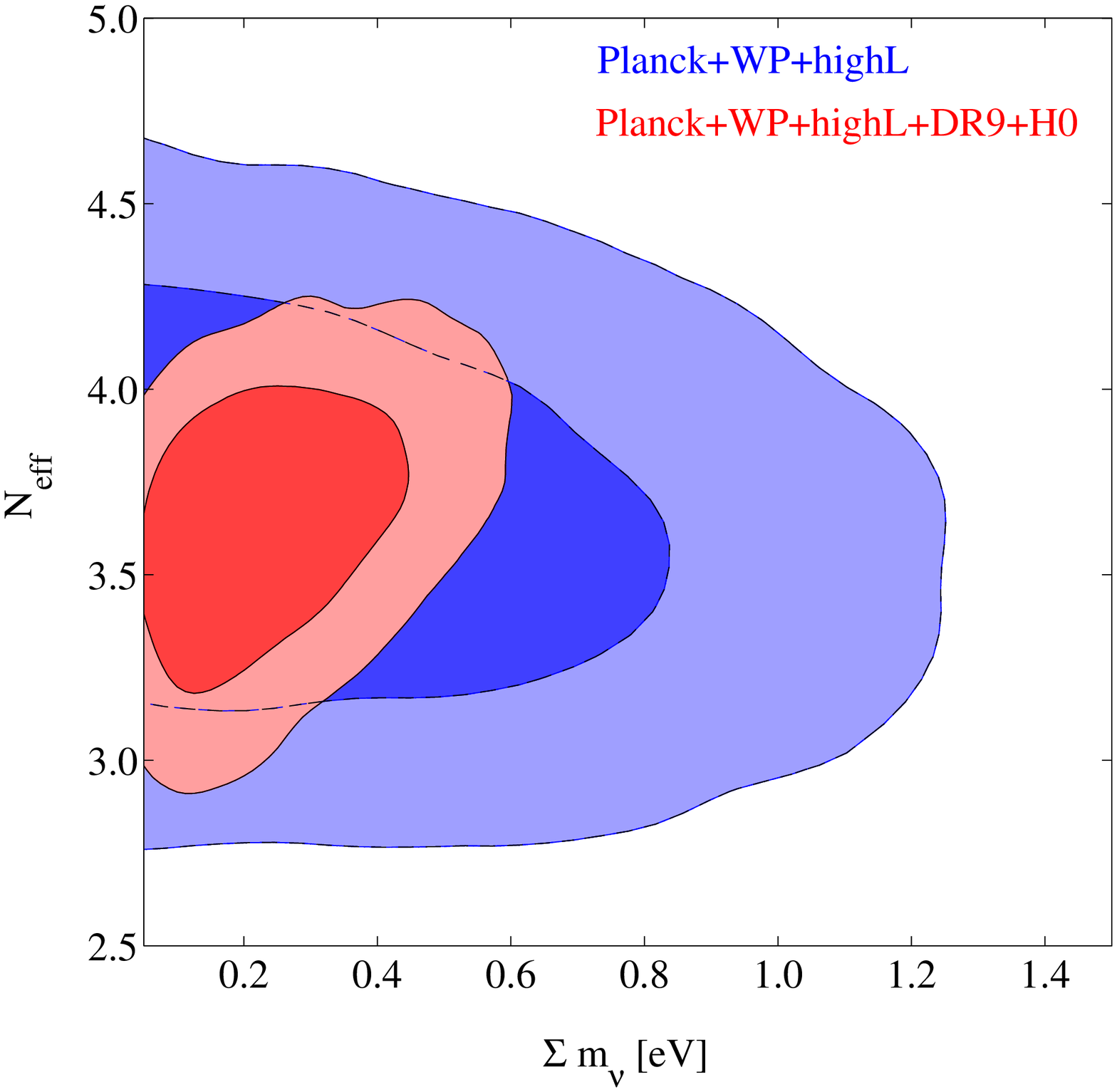}\\
\end{tabular}
 \caption{$68\%$ and $95\%$~c.l. 2D marginalized posterior in the $\summnu - \neff$ plane, blue contours refer to the constraints from the combination of Planck+WP+highL, red contours include also $H_0$ (left panel), DR9 (middle panel) and $H_0$ and DR9 (right panel).}
\label{fig:neffmnutot}
\end{figure*}

Concerning the effects of external non CMB data sets, we include in the analyses of a $\Lambda$CDM model with massive neutrinos and a varying lensing amplitude the $H_0$ prior and the DR9 data. On one hand with the inclusion of the $H_0$ prior we obtain a better constraint on $\neff$, driving $\neff$ from $\neff=3.65\pm0.38$ to $\neff=3.81\pm0.28$ (68\% c.l. errors). So the combination of the data sets Planck+WP+highL+$H_0$ provides a stronger evidence ($2.7\sigma$) for an extra dark radiation component. On the other hand $H_0$ leads to tighter constraints on the 95\% c.l. upper bound of the sum of neutrino masses, moving it from $\summnu<1.03$~eV to $\summnu<0.66$~eV at 95\% c.l. (see Figure \ref{fig:neffmnutot} left panel). The same effect on $\summnu$ can be achieved by including DR9, but in this case $\neff$ remains close to the standard value $\neff=3.26\pm0.30$ (68\% c.l.) (see Figure \ref{fig:neffmnutot} middle panel). The joint effect of adding both a $H_0$ prior and the galaxy clustering information from BOSS DR9 is shown in Figure \ref{fig:neffmnutot} (right panel): the 95\% upper bound on the sum of neutrino masses is tightened both by the prior on $H_0$ and the DR9 galaxy clustering information: an extra dark radiation component is favored at $2.3~\sigma$ level ($\neff=3.65\pm0.26$, 68\% c.l.).

\subsection{Constraints on $\ceff$ and $\cvis$: perturbation parameters}
\label{sec:neffceffcvis}

Table \ref{tab:neffceffcvis} reports the constraints on the perturbation parameters of a varying number of relativistic species. The neutrino perturbation parameters are not strongly affected by the inclusion of the $H_0$ prior: the constraints on $\ceff$ and $\cvis$ remain almost the same. Interestingly both the effective sound speed and the viscosity parameter show a deviation from the standard value $0.33$ being: $\ceff=0.309\pm0.012$ and $\cvis=0.56\pm0.17$ at 68\% c.l. for the basic data set Planck+WP+highL,
consistent with the results of \cite{Gerbino:2013ova}.
Furthermore we can notice that varying the neutrino perturbation parameters does not change our conclusions on the effective number of relativistic species; the bounds on $\neff$ turn out to be almost the same as those reported in Table \ref{tab:neffyhe}: varying $\ceff$ and $\cvis$ we get $\neff=3.40\pm0.34$, while we obtained $\neff=3.44\pm0.35$ with standard $\ceff$ and $\cvis$.
\begin{table*}
\begin{center}
\begin{tabular}{lcc}
\hline
\hline
   &             Planck+WP+highL & Planck+WP+highL \\
     &                                       & +$H_0$                   \\
\hline
\hspace{1mm}\\
$\neff$   & $3.40\pm0.34$ & $3.56\pm0.26$ \\
\hline
\hspace{1mm}\\
$\ceff$   & $0.309\pm0.012$ & $0.310\pm0.012$\\
\hspace{1mm}\\
$\cvis$   & $0.56\pm0.17$ & $0.57\pm0.17$  \\
\hline
\hline
\end{tabular}
\caption{Marginalized 68\% constraints on $\neff$, $\ceff$ and $\cvis$.}
\label{tab:neffceffcvis}
\end{center}
\end{table*}

\section{Conclusions}
\label{sec:conclusions}

The newly released Planck data have provided us with an extremely precise picture of the cosmic microwave background, confirming the standard $\Lambda$CDM model. However the exact properties of the dark sector are still under discussion and, in particular,  there is no strong argument against the existence of a dark radiation component. On the contrary, combining CMB data with measurements of galaxy clustering and of the Hubble constant leads to an evidence for a non standard number of relativistic species.

In this article we have illustrated the effects of an additional relativistic component on the temperature power spectrum and we have reviewed the most promising models to explain the presence of this component: sterile neutrinos, axions, decay of massive particles, interactions between dark matter and dark radiation sectors.

We have focused on the hypothesis of a link between cosmology and neutrino physics that can explain the cosmologically inferred excess in the number of relativistic species in terms of sterile neutrinos whose existence could explain some short baseline neutrino oscillations results.
In this framework we have updated the cosmological constraints on massive neutrinos including the new Planck CMB data and the matter power spectrum from BOSS DR9. Including also a prior on the Hubble constant from the Hubble Space Telescope measurements, our results show a preference for a non standard number of neutrino species at $2.3~\sigma$ with $\neff=3.65\pm0.26$ at 68\% c.l. and an upper bound on the sum of neutrino masses of $0.51$~eV at 95\% c.l..

However the relevance of these cosmological constraints on dark radiation depends on the model and on the data sets.

We have stressed the impact of the lensing amplitude on these results: the inclusion of a varying lensing amplitude drives the results towards a more statistically significant detection of dark radiation.

Concerning the data sets, the $H_0$ prior also leads to a better constraint on $\neff$. The former effect is related to the $2.5~\sigma$ tension among Planck and HST measurements of the Hubble constant that must be fixed (\cite{Marra:2013rba}).

Finally our results confirm a significant deviation from the standard values ($\ceff=1/3$ and $\cvis=1/3$) expected for a free streaming dark radiation component. We find $\ceff=0.309\pm0.012$ and $\cvis=0.56\pm0.17$ at 68\% c.l., allowing for further consideration on the nature of dark radiation.

In conclusion, there is still ample room for interesting new discoveries of physics beyond the standard model in the form of dark radiation.

\section{Acknowledgments}
We acknowledge the European ITN project Invisibles (FP7-PEOPLE-2011-ITN, PITN-GA-2011-289442-INVISIBLES).

\end{document}